\documentclass[prd,aps,nofootinbib,onecolumn,showpacs,12pt]{revtex4-1}
\usepackage{graphicx,epsfig}

\usepackage{epsf}
\usepackage{graphicx}
\def\ds{\displaystyle}

\def\ds{\displaystyle}
\def\beq{\begin{equation}}
\def\eeq{\end{equation}}
\def\bea{\begin{eqnarray}}
\def\eea{\end{eqnarray}}
\def\beeq{\begin{eqnarray}}
\def\eeeq{\end{eqnarray}}

\def\vel{\left|}
\def\ver{\right|}
\def\nnb{\nonumber}
\def\ga{\left(}
\def\dr{\right)}

\def\rar{\rightarrow}
\def\nnb{\nonumber}

\def\ba{\begin{array}}
\def\ea{\end{array}}

\def\xis0{{\Xi^{*0}}}

\def\g5{\gamma_5}

\def\es{\!\!\! &=& \!\!\!}

\begin{document}

\title{\bf Systematic  analysis of the $ B_s\rightarrow f_0 \ell^+\ell^- $ in the universal extra dimension }
\author{ V. Bashiry$^{\dag1}$, K. Azizi$^{\ddag2}$ \\
$^{\dag}$ Cyprus International University, Faculty of Engineering, Department of Computer Engineering,
Nicosia, Northern Cyprus,  Mersin 10, Turkey\\
$^{\ddag}$ Department of Physics, Do\u gu\c s University,
Ac{\i}badem-Kad{\i}k\"oy, 34722 Istanbul, Turkey\\
$^1$bashiry@ciu.edu.tr\\
$^2$kazizi@dogus.edu.tr}

\begin{abstract}
Using form factors enrolled to the transition matrix elements and calculated via light-cone QCD sum rules including next-to-leading order corrections
 in the strong coupling constant, we provide a systematic analysis of the  $ B_s\rightarrow f_0 (980)\ell^+\ell^- $ both in 
the standard  and universal extra dimension models. In particular, we discuss sensitivity of the differential branching ratio and various double lepton polarization asymmetries
on the compactification factor of extra dimension and show how the results of the extra dimension model deviate from the standard model predictions. The order of branching ratio makes this decay mode 
possible to be checked
at LHCb in near future.
\end{abstract}

\maketitle

\section{Introduction}
It is well known that the decays of the $ B_s$ meson are very promising tools to constrain the standard model (SM) parameters, serve to determine  the elements of the Cabibbo-Kobayashi-Maskawa (CKM) matrix,
enable us to understand  the origin of the CP violation and help us search for new physics (NP) effects beyond the SM. One of the possible decay modes of the $ B_s$ meson is the  semileptonic $ B_s\rightarrow f_0 (980)\ell^+\ell^- $. This channel proceeds via flavor-changing
neutral currents (FCNC) transition of  $b\rightarrow s \ell^+\ell^- $ at quark level, which is induced  at loop level in
the SM and is therefore sensitive to NP effects. The extra dimensions as NP effects can contribute to such loop level transitions and enhance the branching ratio.
It has been shown that, in the presence of a single
universal extra dimension (UED) compactified on a circle with radius $R$, the branching ratios of the $B_s\rightarrow \eta^{(')}\ell^+\ell^-/ \nu \bar\nu $ and $B_s\rightarrow \phi\nu \bar\nu $ which are also based on
$b\rightarrow s \ell^+\ell^-/\nu \bar\nu $, increase significantly
at lower values of the compactification factor, $1/R$ \cite{Colangelo1,Colangelo2}.

In the present work, we investigate the effect of the UED on some physical observables related to the semileptonic  $ B_s\rightarrow f_0 (980)\ell^+\ell^- $. The UED with a single extra dimension
 called the  Appelquist, Cheng and Dobrescu (ACD) model \cite{acd},
 is a kind of extra dimension (ED) \cite{antoniadis1,antoniadis2,arkani} which allows the SM fields (both gauge bosons and fermions) to propagate in the extra dimensions (for more details see for instance \cite{Hooper}). The ACD model has been previously applied to 
many decay channels. For some of them see \cite{R7624,wangying,R7601,wangying,fazio,sirvanli,kank1,yee,bashiry,aliev,ahmet,nihan,cog,Mohanta,Haisch:2007vb} and references therein.

 In the SM, the effective Hamiltonian
 describing the $ B_s\rightarrow f_0 (980)\ell^+\ell^- $ transition at quark level can be written as:
 \begin{eqnarray} \label{e8401}
{\cal H}^{eff} &=& {G_F \alpha_{em} V_{tb}
V_{ts}^\ast \over 2\sqrt{2} \pi} \Bigg[ C_9^{eff}
\bar{s}\gamma_\mu (1-\gamma_5) b \, \bar{\ell} \gamma^\mu
\ell + C_{10} \bar{s} \gamma_\mu (1-\gamma_5) b \,
\bar{\ell} \gamma^\mu
\gamma_5 \ell \nonumber \\
&-&  2 m_b C_7^{eff}  {1\over q^2}\bar{s} i
\sigma_{\mu\nu}q^{\nu} (1+\gamma_5) b \, \bar{\ell} \gamma^\mu
\ell \Bigg]~.
\end{eqnarray}
 where $V_{ij}$ are elements of
the CKM matrix,  $\alpha_{em}$  is the fine structure
constant, $G_F$ is the Fermi  constant, and $C_7^{eff}$, $C_9^{eff}$ and $C_{10}$  are
Wilson coefficients.  In the ACD model, the form of the effective Hamiltonian remains unchanged but due to the interaction of the Kaluza-Klein (KK)
particles  with the usual SM particles and also  with themselves, the Wilson coefficients are modified. This modification is done in
\cite{Buras:2002ej,R7623,R7626,R7627,R762777} in
leading logarithmic approximation in a way that
each Wilson coefficient is written in terms of  some periodic
functions as:
\begin{equation}
 F(x_t,1/R)=F_0(x_t)+\sum_{n=1}^{\infty}F_n(x_t,x_n),
\label{functions}
\end{equation}
 where,  $F_0 (x_t )$ is ordinary SM part   and the other part can be written in terms of the
compactification factor $1/R$, by means of the following definitions:
\begin{equation}
 x_{t}=m_{t}^{2}/M_{W}^{2},~~~~~~~~~~~~x_n=m_n^2/m_W^2,~~~~~~~~~~~~~~m_n=n/ R,
\end{equation}
  where, $m_{t}$ is the mass of the top quark, $M_{W}$ is the mass of the $W$ boson, $m_n$ is the  mass of the KK particles and $n=0$ corresponds to the ordinary SM  particles. 
Few comments about the lower bound of the compactification factor are in order.  From the electroweak precision tests, the lower limit for $1/R$ had been previously obtained as $%
250~GeV$ in \cite{acd,yee} if $M_h\geq250~GeV$ expressing larger KK contributions to the low energy FCNC processes,
 and $300~GeV$  if $M_h\leq250~GeV$. 
Analysis of the $B\rightarrow X_s\gamma$ transition  and also anomalous magnetic moment
  had shown also that the experimental data are in a good agreement with the ACD model if $1/R\geq 300~GeV$ \cite{ikiuc}. Taking into account the leading order
(LO) contributions due to the exchange of KK modes as
well as the available next-to-next-to-leading order (NNLO) corrections 
to also $B(B \rightarrow X_s \gamma)$ in the SM, the authors of \cite{Haisch:2007vb} have obtained a lower bound on the inverse compactification radius $600~GeV$. Using the electroweak precision measurements and also
some cosmological constraints, the authors of \cite{Gogoladze:2006br} and \cite{Cembranos:2006gt} have found that the lower limit on compactification factor is in or above the $500~GeV$ range. We will plot the physical observables
under consideration in the range $1/R\in[200-1000]GeV$ just to clearly show how the results of the UED deviate from those of the SM and grow decreasing the $1/R$ .

 The numerical values for different Wilson coefficients  both in the SM and ACD model in the
range $1/R\in[200-1000]GeV$  are presented in Table~\ref{SCXY}.
From this Table, we see that in the ACD model, the $C_{10}$ is enhanced and
$C_7^{eff}$ is suppressed considerably in comparison with their values in the SM.
\begin{table}[hbt]
\begin{center}
\begin{tabular}{|c||c|c|c|}\hline
  $1/R~[{\rm GeV}]$   & $C_7^{eff}$ & $C_{10}$ & $C_9^{eff}$
 \\\hline
$ 200$  &
 $-0.195212$ & $-5.61658$ &
 $4.83239 + 3.59874i$\\\hline
$400$  & $ -0.266419$ & $-4.65118$ & $4.7538 + 3.54366i$\\\hline
 $600$  & $-0.283593$ & $-4.43995$ & $ 4.7366 + 3.53161i$\\\hline
  $800$  & $-0.29003$ &
$-4.36279$ & $4.73032 + 3.52721i$\\\hline
  $1000$  & $-0.293092$ & $-4.32646$ & $4.72736 + 3.52514i $\\\hline
 SM      &
$-0.298672$ & $-4.26087 $ &
 $4.72202 + 3.52139i$ \\\hline
\end{tabular}
\end{center}
\caption[]{\small Numerical values for $C_7^{eff}$,   $C_{10}$ and
values of $C_9^{eff}$ at transferred momentum square, $q^2=14$   for different values of $1/R$ as well as the
SM \label{SCXY}}
\end{table}

Here we should mention that, besides the aforementioned contributions,  the  Wilson coefficient $C_9^{eff}$ receives also long distance contributions from  $J/\psi$ family
 parameterized  using Breit--Wigner ansatz \cite{R8401}, i.e., 
\bea Y_{\rm LD} = {3\pi \over \alpha_{em}^2} C^{(0)} \sum_{V_i=\psi(1s)
\cdots \psi(6s)} \ae_i \, {\Gamma(V_i \rar \ell^+ \ell^-) m_{V_i}
\over m_{V_i}^2- q^2-im_{V_i}\Gamma_{V_i}} ~, \nnb \eea where, $C^{(0)}=0.362$. As the  phenomenological factors, $\ae_i$ have not known for the transition under 
consideration, we will choose the values of the $q^2$ which do not lie on the $J/\psi$ family resonances. This is possible in the case of the differential branching ratio and double--lepton
 polarization asymmetries under consideration in this work, however, to calculate the total branching ratio as well as the average double--lepton polarization asymmetries,
 which require integration over $q^2$, one should take also into account such contributions.  For more details about the long distance 
contributions see for instance \cite{cana1,cana2}.

The layout of the paper is as follows.  In the next section, we present the transition matrix elements expressed in terms of form factors and the formula for decay rate as well as various double lepton polarization asymmetries.
In the last section, we numerically analyze the observables in terms of compactification factor, $1/R$ of the ACD model and discuss the results.

\section{Transition matrix elements and observables related to the  $B_s\rightarrow f_0 (980)\ell^+\ell^-$ channel }

To find the amplitude of the  decay channel in question, we need to sandwich the aforementioned effective Hamiltonian between the initial and final states. As a result, we obtain the following transition matrix elements defined
in terms of form factors $F_0(q^2)$, $F_1(q^2)$ and $F_T(q^2)$:
\begin{eqnarray}
&&\langle f_0(p_{f_0})|\bar s \gamma_\mu\gamma_5 b |\overline {B}_s(p_{B_s})\rangle = -i \Big\{F_1(q^2)\Big[P_\mu
 -\frac{m_{B_s}^2-m_{f_0}^2}{q^2}q_\mu\Big] + F_0(q^2)\frac{m_{B_s}^2-m_{f_0}^2}{q^2}q_\mu\Big\}, \,\,\,
 \label{F1-F0}
 \end{eqnarray}
  \begin{eqnarray}
&&\langle {f_0}(p_{f_0})|\bar s\sigma_{\mu\nu}\gamma_5q^\nu b |\overline {B}_s(p_{B_s})\rangle = -\frac{F_T(q^2)}{m_{B_s}+m_{f_0}}   \Big[q^2P_\mu
 -(m_{B_s}^2-m_{f_0}^2)q_\mu\Big], \label{FT}
\end{eqnarray}
where $P=p_{B_s}+p_{f_0}$ and $q=p_{B_s}-p_{f_0}$. For simplicity in some parts of calculations, it is
 convenient to introduce the auxiliary form factors $f_+$ and $f_-$,
 \begin{eqnarray}
 \langle {f_0}(p_{f_0})|\bar s\gamma_\mu\gamma_5 b |\overline {B}_s(p_{B_s})\rangle& =&
 -i\left\{f_+(q^2) P_\mu +f_-(q^2)q_\mu\right\} \nonumber \\
 \end{eqnarray}
 such that
\begin{eqnarray}
 F_1(q^2)&=& f_+(q^2) \,\, , \nonumber \\
 F_0(q^2)&=&f_+(q^2)+\frac{q^2}{m_{B_s}^2-m_{f_0}^2} f_-(q^2) \,\, .
\end{eqnarray}
The form factors,  $F_1$, $F_0$ and $F_T$  are calculated via light-cone QCD sum rules both  at the leading order and the next-to-leading order corrections in \cite{colangelo3}.
 We use the latter to analyze the considered physical observables. The fit parametrization of the form factors including next-to-leading order corrections in $\alpha_s$ is given as \cite{colangelo3}:
\begin{eqnarray}
 F_i(q^2)&=&\frac{F_i(0)}{1-a_iq^2/m_{B_s}^2+b_i(q^2/m_{B_s}^2)^2}\,\,,\label{eq:formfactors-fit-form}
\end{eqnarray}
where $F_i$ denotes any function among $F_{1,0,T}$.  The parameters $a_i$ and $b_i$ as well as $F_i(0)$ are given in Table \ref{tab:formfactors-mass-zero}.
\begin{table}
\caption{ ${ B_s \to  f_0(980)}$ transition form factors  including  next-to-leading order corrections \cite{colangelo3}.
}\label{tab:formfactors-mass-zero}
\begin{tabular}{c c c c c }
\hline\hline
& $F_i(q^2=0)$  & $a_i$ & $b_i$   \\\hline
$F_1 $   &   $0.238\pm0.036$      &   $1.50^{+0.13}_{-0.09}$  & $0.58^{+0.09}_{-0.07}$ \\
$F_0 $   &   $ 0.238\pm 0.036$&$0.53^{+0.14}_{-0.10}$  &   $-0.36^{+ 0.09}_{-0.08}$    \\
$F_T$   &   $ 0.308\pm0.049$&   $1.46^{+0.14}_{-0.10}$  & $0.58^{+ 0.09}_{-0.07}$    \\
\hline\hline
\end{tabular} \end{table}
%

Now we proceed to calculate some observables such as differential decay rate and double lepton polarization asymmetries. With the matrix elements in terms of form factors one can easily obtain
the $1/R$-dependent differential decay rate
as:
 \begin{eqnarray}
\frac{d\Gamma(\bar B_s\to f_0\ell^+\ell^-)}{dq^2}(q^2,1/R) &&=  \frac{G_F^2 \alpha^2_{em} |V_{tb}|^2|V^*_{ts}|^2 \sqrt{\lambda}}{512 m_{B_s}^3\pi^5}
 \frac{v}{3q^2} \Bigg[6m_\ell^2 |C_{10}(1/R)|^2(m_{B_s}^2-m_{f_0}^2)^2F_0^2(q^2) \nonumber\\
 &&+(q^2+2m_\ell^2)\lambda\bigg| C_9^{eff}(q^2,1/R)F_1(q^2)+\frac{2C_7^{eff}(1/R)(m_b-m_s) F_T(q^2)}{m_{B_s}+m_{f_0}}\bigg|^2  \nonumber\\
 &&+ |C_{10}(1/R)|^2(q^2-4m_\ell^2)\lambda F_1^2(q^2)
\Bigg],\label{eq:partial-decay}
 \end{eqnarray}
 with $v=\sqrt{1-\frac{4m_\ell^2}{q^2}}$,
$\lambda=\lambda(m_{B_s}^2,m_{f_0}^2,q^2)$ with $\lambda(a,b,c)=(a-b-c)^2-4bc$ and $m_\ell$ is the lepton's mass.

To calculate the double--polarization
asymmetries,  we consider the  polarizations of both lepton and anti-lepton, 
simultaneously and introduce the following  spin projection
operators for the lepton $\ell^-$ and the anti-lepton $\ell^+$ (see also  \cite{bas1,bas2,Fukae}): \bea
\Lambda_1 ~\es~ \frac{1}{2} (1+\gamma_5\!\!\not\!{s}_i^-)~,\nnb \\
\Lambda_2 ~\es~ \frac{1}{2} (1+\gamma_5\!\!\not\!{s}_i^+)~,  \eea
 where $i=L,N$ and $T$
correspond to the longitudinal, normal and transversal
polarizations, respectively. Now, we   define the following  orthogonal
vectors $s^\mu$ in the rest frame of lepton and anti-lepton: \bea \label{e6010} s^{-\mu}_L
~\es~ \ga 0,\vec{e}_L^{\,-}\dr =
\ga 0,\frac{\vec{p}_-}{\vel\vec{p}_- \ver}\dr~, \nnb \\
s^{-\mu}_N ~\es~ \ga 0,\vec{e}_N^{\,-}\dr = \ga
0,\frac{\vec{p}_{f_0}\times
\vec{p}_-}{\vel \vec{p}_{f_0}\times \vec{p}_- \ver}\dr~, \nnb \\
s^{-\mu}_T ~\es~ \ga 0,\vec{e}_T^{\,-}\dr = \ga 0,\vec{e}_N^{\,-}
\times \vec{e}_L^{\,-} \dr~, \nnb \\
s^{+\mu}_L ~\es~ \ga 0,\vec{e}_L^{\,+}\dr =
\ga 0,\frac{\vec{p}_+}{\vel\vec{p}_+ \ver}\dr~, \nnb \\
s^{+\mu}_N ~\es~ \ga 0,\vec{e}_N^{\,+}\dr = \ga
0,\frac{\vec{p}_{f_0}\times
\vec{p}_+}{\vel \vec{p}_{f_0}\times \vec{p}_+ \ver}\dr~, \nnb \\
s^{+\mu}_T ~\es~ \ga 0,\vec{e}_T^{\,+}\dr = \ga 0,\vec{e}_N^{\,+}
\times \vec{e}_L^{\,+}\dr~, \eea where $\vec{p}_\mp$  are the three--momenta of the leptons $\ell^\mp$ and
$\vec{p}_{f_0}$ is three-momentum of the final $f_0$
meson in the center of mass  (CM) frame  of $\ell^- \,\ell^+$. By Lorenz transformations, the longitudinal unit vectors are boosted to the CM frame of $\ell^-
\ell^+$, \bea \label{e6011} \ga s^{-\mu}_L
\dr_{CM} ~\es ~\ga \frac{\vel\vec{p}_- \ver}{m_\ell}~,
\frac{E \vec{p}_-}{m_\ell \vel\vec{p}_- \ver}\dr~,\nnb \\
\ga s^{+\mu}_L \dr_{CM} ~\es ~\ga \frac{\vel\vec{p}_-
\ver}{m_\ell}~, -\frac{E \vec{p}_-}{m_\ell \vel\vec{p}_-
\ver}\dr~, \eea while the other two vectors are kept the same. Now, we  define the double--lepton polarization asymmetries as
 \cite{bas1,bas2,Fukae}: \bea \label{e6012} P_{ij}(\hat{s}) =
\frac{\ds{\Bigg(
\frac{d\Gamma}{d\hat{s}}(\vec{s}_i^-,\vec{s}_j^+)}-
\ds{\frac{d\Gamma}{d\hat{s}}(-\vec{s}_i^-,\vec{s}_j^+) \Bigg)} -
\ds{\Bigg( \frac{d\Gamma}{d\hat{s}}(\vec{s}_i^-,-\vec{s}_j^+)} -
\ds{\frac{d\Gamma}{d\hat{s}}(-\vec{s}_i^-,-\vec{s}_j^+)\Bigg)}}
{\ds{\Bigg( \frac{d\Gamma}{d\hat{s}}(\vec{s}_i^-,\vec{s}_j^+)} +
\ds{\frac{d\Gamma}{d\hat{s}}(-\vec{s}_i^-,\vec{s}_j^+) \Bigg)} +
\ds{\Bigg( \frac{d\Gamma}{d\hat{s}}(\vec{s}_i^-,-\vec{s}_j^+)} +
\ds{\frac{d\Gamma}{d\hat{s}}(-\vec{s}_i^-,-\vec{s}_j^+)\Bigg)}}~,
\eea where the subindex $j$ also stands for the $L,~N$ or $T$ polarization. The  subindexses, $i$ and $j$ correspond
to the lepton and anti-lepton, respectively.
 Using the above  definitions, the various $1/R$-dependent double lepton polarization asymmetries are obtained in the following way:
\begin{eqnarray}
P_{LL}(\hat{s},1/R)&=&\frac{-4m_{B_{s}}^2}{3\Delta(\hat{s},1/R)}Re[-24m_{B_{s}}^2\hat{m}_{l}^2(1-\hat{r}_{f_{0}})C^{*}D+\lambda^{'}
m_{B_{s}}^2(1+v^2)|A|^2\nonumber\\&-& 12m_{B_{s}}^2 \hat{m}_{l}^2\hat{s}|D|^2
+m_{B_{s}}^2|C|^2(2\lambda^{'}-(1-v^2)(2\lambda^{'}+3(1-\hat{r}_{f_{0}})^2))],\\
P_{LN}(\hat{s},1/R)&=&\frac{-4\pi m_{B_{s}}^{3} \sqrt{\lambda^{'}
\hat{s}}}{\hat{s}\Delta(\hat{s},1/R)}
Im[-m_{B_{s}}\hat{m}_{l}\hat{s}A^{*}D -m_{B_{s}}\hat{m}_{l}(1-\hat{r}_{f_{0}})A^{*}C],\\
P_{NL}(\hat{s},1/R)&=&-P_{LN}(\hat{s},1/R),\\
P_{LT}(\hat{s},1/R)&=&\frac{4\pi
m_{B_{s}}^3\sqrt{\lambda^{'}\hat{s}}}{\hat{s}\Delta(\hat{s},1/R)}Re[m_{B_{s}}\hat{m}_{l}v(1-\hat{r}_{f_{0}})|C|^2+m_{B_{s}}\hat{m}_{l}v\hat{s}C^{*}D],\\
P_{TL}(\hat{s},1/R)&=&P_{LT}(\hat{s},1/R),\\
P_{NT}(\hat{s},1/R)&=&-\frac{8m_{B_{s}}^2v}{3\Delta(\hat{s},1/R)}Im[2\lambda^{'} m_{B_{s}}^2A^{*}C],\\
P_{TN}(\hat{s},1/R)&=&-P_{NT}(\hat{s},1/R),\\
P_{TT}(\hat{s},1/R)&=&\nonumber\frac{4m_{B_{s}}^2}{3\Delta(\hat{s},1/R)}Re[-24
m_{B_{s}}^2\hat{m}_{l}^2(1-\hat{r}_{f_{0}})C^{*}D-\lambda^{'}
m_{B_{s}}^2(1+v^2)|A|^2-12m_{B_{s}}^2\hat{m}_{l}^2\hat{s}|D|^2\\&+&m_{B_{s}}^2|C|^2\{2\lambda^{'}-(1-v^2)(2\lambda^{'}+3(1-\hat{r}_{f_{0}})^2)\}],\\
P_{NN}(\hat{s},1/R)&=&\nonumber\frac{4m_{B_{s}}^2}{3\Delta(\hat{s},1/R)}Re[24
m_{B_{s}}^2\hat{m}_{l}^2(1-\hat{r}_{f_{0}})C^{*}D-\lambda^{'}
m_{B_{s}}^2(3-v^2)|A|^2+12m_{B_{s}}^2\hat{m}_{l}^2\hat{s}|D|^2\\&+&m_{B_{s}}^2|C|^2\{2\lambda^{'}-(1-v^2)(2\lambda^{'}-3(1-\hat{r}_{f_{0}})^2)\}],
\end{eqnarray}
where, $\hat{s}=\frac{q^2}{m_{B_{s}}^2} $,
 $\hat{r}_{f_{0}}=\frac{m^2_{f_{0}}}{m_{B_{s}}^2} $, $\hat{m}_{l}=\frac{m_l}{m_{B_{s}}}$, $\lambda^{'}=\lambda(1,\hat{r}_{f_{0}},\hat{s})$ and
 \begin{eqnarray}
\Delta(\hat{s},1/R)&=&\nonumber\frac{4m_{B_{s}}^2}{3}Re[24
m_{B_{s}}^2\hat{m}_{l}^2(1-\hat{r}_{f_{0}})D^{*}C+\lambda^{'}
m_{B_{s}}^2(3-v^2)|A|^2+12m_{B_{s}}^2\hat{m}_{l}^2\hat{s}|D|^2\\
&+&
m_{B_{s}}^2|C|^2\{2\lambda^{'}-(1-v^2)(2\lambda^{'}-3(1-\hat{r}_{f_{0}})^2)\}],
\end{eqnarray}
with
\begin{eqnarray}
A=A(\hat{s},1/R) &=&2C_{9}^{eff} (\hat{s},1/R) f_+(\hat{s})
-
4C_{7}^{eff}(1/R)(m_b+m_s) \frac{F_T(\hat{s})}{m_{B_{s}}+m_{f_{0}}} ,
 \nonumber \\
B=B(\hat{s},1/R) &=& 2C_{9}^{eff} (\hat{s},1/R) f_-(\hat{s}) +4C_{7}^{eff}(1/R)(m_b+m_s) \frac{F_T(\hat{s})}{(m_{B_{s}}+m_{f_{0}})\hat{s}m_{B_{s}}^2}(m_{B_{s}}^2-m_{f_{0}}^2) ,
 \nonumber \\
C=C(\hat{s},1/R) &=& 2C_{10}(1/R) f_+(\hat{s}) ,
 \nonumber \\
D=D(\hat{s},1/R) &=& 2C_{10}(1/R)f_-(\hat{s}) ~.
 \end{eqnarray}

\section{Numerical Results}
In this section, we numerically analyze the physical observables and discuss their sensitivity to the compactification factor of extra dimension. The main input parameters are form factors in the matrix elements
whose fit parametrization are presented in the previous section. To proceed in numerical calculations, we also need to know the numerical values of the other input parameters. We use the values:
$m_t=167~GeV$, $m_W=80.4~GeV$,
$m_b=4.8~GeV$, $m_s=0.14~GeV$, $m_{\mu}=0.105~GeV$,  $m_{\tau}=1.778$,
 $| V_{tb}V_{ts}^\ast|=0.041$, $G_F = 1.166
\times 10^{-5}~ GeV^{-2}$, $\alpha_{em}=\frac{1}{137}$,
$\tau_{B_s}=1.42\times 10^{-12}~s$, $m_{f_0} =
0.980~GeV$ and $m_{B_{s}} = 5.36~ GeV$.

Considering the central values of the form factors, we plot the dependence of the differential branching ratio and various double lepton polarization asymmetries for the  $ B_s\rightarrow f_0 \ell^+\ell^- $  decay channel on the compactification
factor ($1/R)$ of the extra dimension in figures 1-8. As the results of the electron channel
are very close to those of the $\mu$, we will depict only the results of the $\mu$ and $\tau$ channels. As it is evident from the formulas in the previous section that the observables depend on $\hat s$,  we
will present our results at three fixed values of this parameter for the  $\mu$ and two fixed values  for $\tau$ channel in the allowed kinematical region 
($4\hat{m}_\ell^2\leq \hat{s}\leq (1-\sqrt{\hat{r}_{f_0}})^2$). Note that in each figure we see graphs of the lines with the same colors. The straight line in 
each case depicts the result of the SM and the curve line stands for the ACD model prediction.

\begin{figure}[h!]
\begin{center}
\includegraphics[width=8cm]{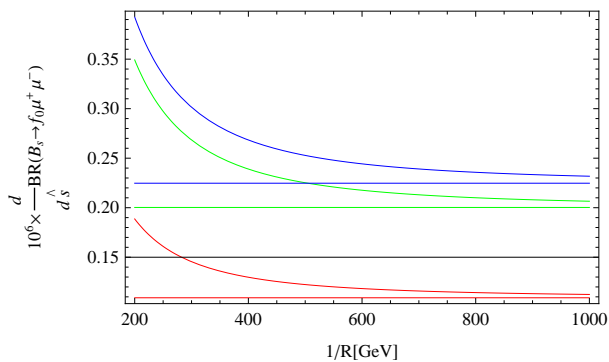}
\end{center}
\caption{Dependence of the branching ratio on the $1/R$ for muon channel  at three fixed
values of the $\hat{s}$. The blue, green and red lines belong to the values $\hat{s}=0.2$, $\hat{s}=0.3$ and $\hat{s}=0.5$, respectively. The
 straight line shows the result of  SM and the curve depicts the ACD model prediction in each case.} \label{fig2}
\end{figure}

\begin{figure}[h!]
\begin{center}
\includegraphics[width=8cm]{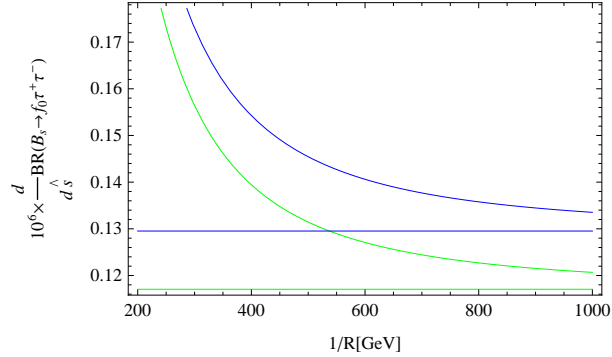}
\end{center}
\caption{Dependence of the branching ratio on the $1/R$ for tau channel  at two fixed
values of the $\hat{s}$. The blue and green  lines belong to the values $\hat{s}=0.5$ and $\hat{s}=0.6$, respectively. The
 straight line shows the result of  SM and the curve depicts the ACD model prediction in each case.} \label{fig4}
\end{figure}
\begin{figure}[h!]
\begin{center}
\includegraphics[width=8cm]{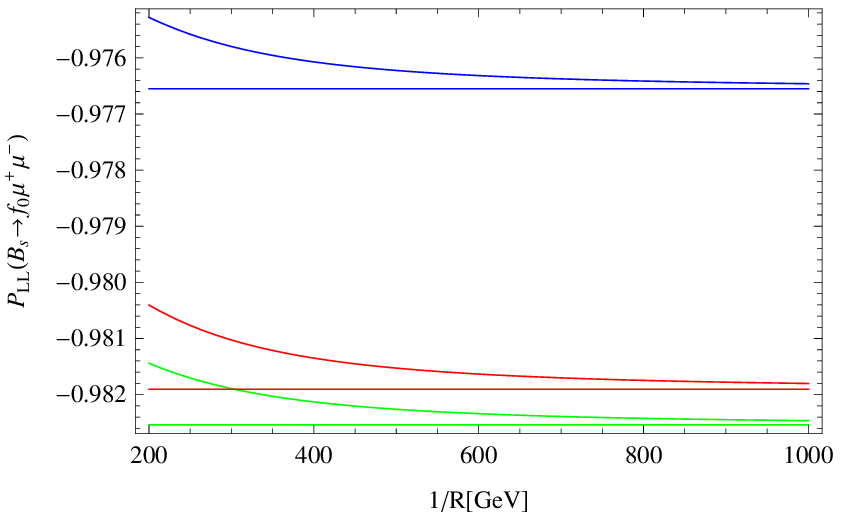}
\includegraphics[width=8cm]{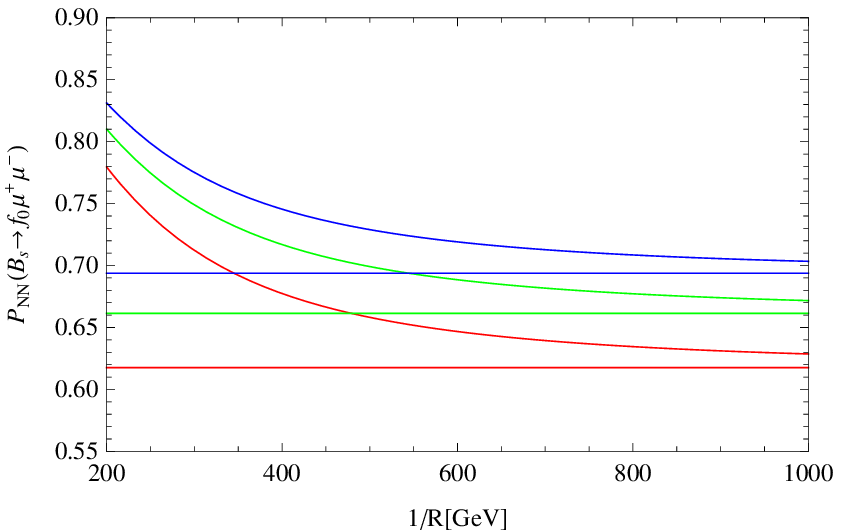}
\end{center}
\caption{Dependence of the $P_{LL}$ and $P_{NN}$ on the $1/R$ for muon channel  at three fixed
values of the $\hat{s}$. The blue, green and red lines belong to the values $\hat{s}=0.2$, $\hat{s}=0.3$ and $\hat{s}=0.5$, respectively. The
 straight line shows the result of  SM and the curve depicts the ACD model prediction in each case.} \label{fig6}
\end{figure}

\begin{figure}[h!]
\begin{center}
\includegraphics[width=8cm]{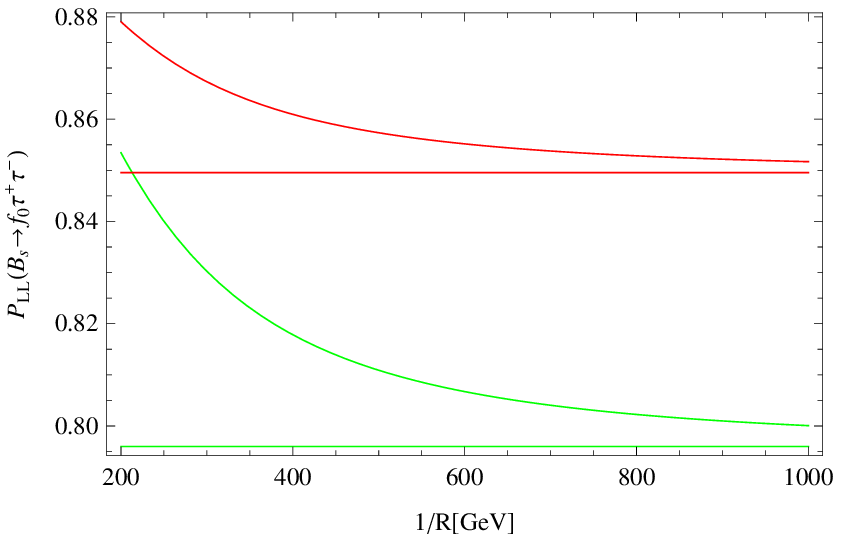}
\includegraphics[width=8cm]{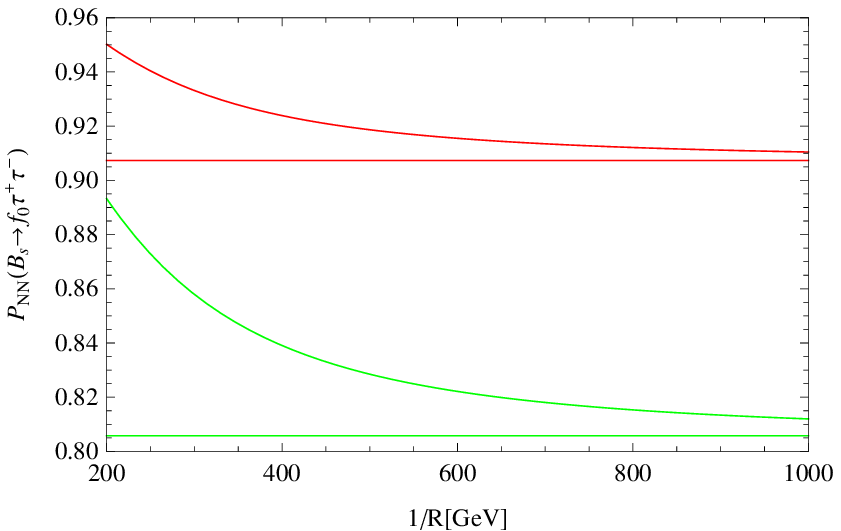}
\end{center}
\caption{Dependence of the $P_{LL}$ and $P_{NN}$ on the $1/R$ for tau channel  at two fixed
values of the $\hat{s}$. The green and red lines belong to the values $\hat{s}=0.5$ and $\hat{s}=0.6$, respectively. The
 straight line shows the result of  SM and the curve depicts the ACD model prediction in each case.} \label{fig8}
\end{figure}


\begin{figure}[h!]
\begin{center}
\includegraphics[width=8cm]{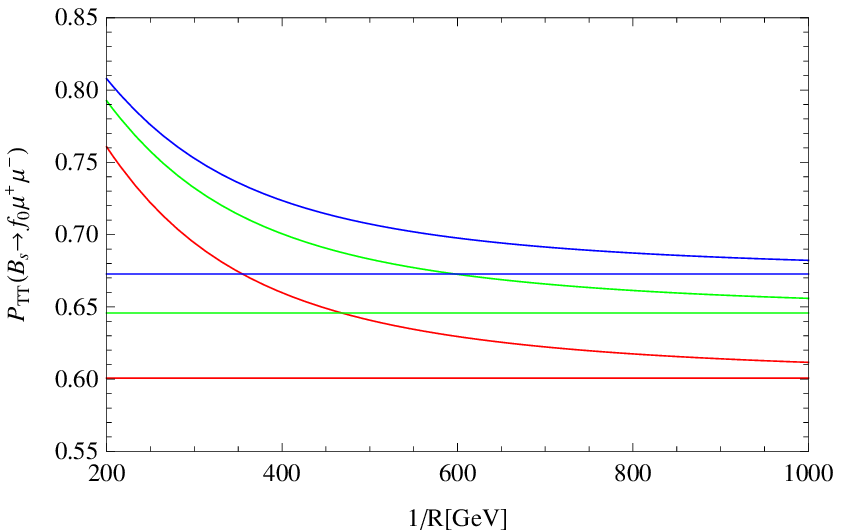}
\includegraphics[width=8cm]{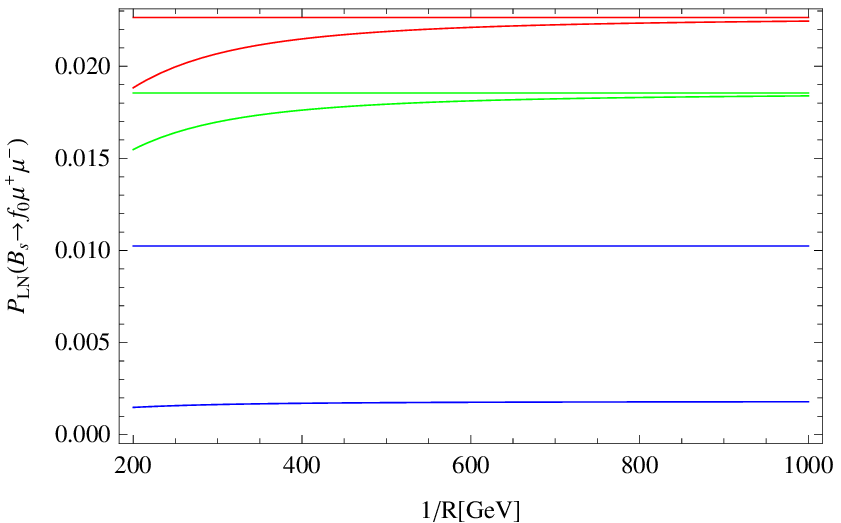}
\end{center}
\caption{The same as FIG. 3 but for $P_{TT}$ and $P_{LN}$.} \label{fig14}
\end{figure}

\begin{figure}[h!]
\begin{center}
\includegraphics[width=8cm]{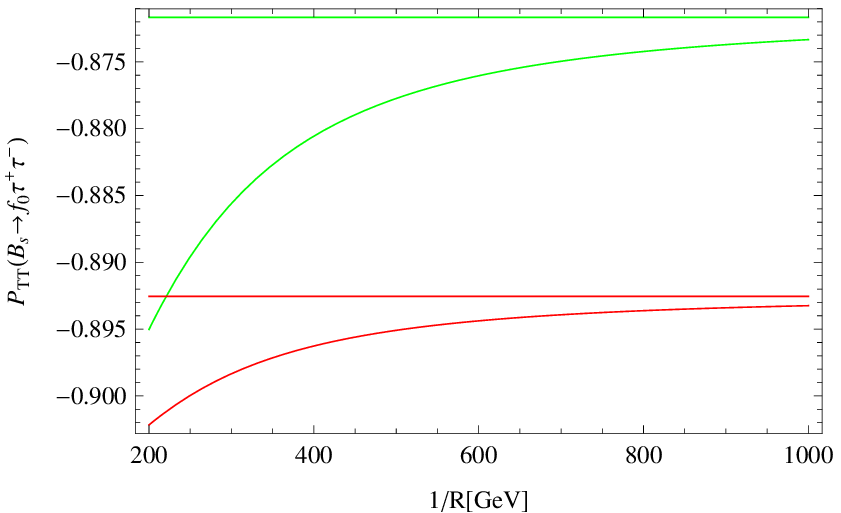}
\includegraphics[width=8cm]{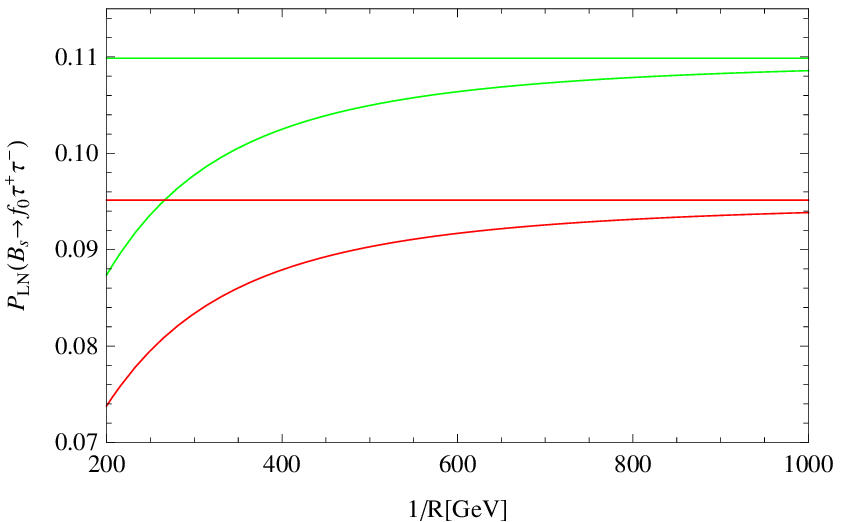}
\end{center}
\caption{The same as FIG. 4 but for $P_{TT}$ and $P_{LN}$.} \label{fig16}
\end{figure}

%

\begin{figure}[h!]
\begin{center}
\includegraphics[width=8cm]{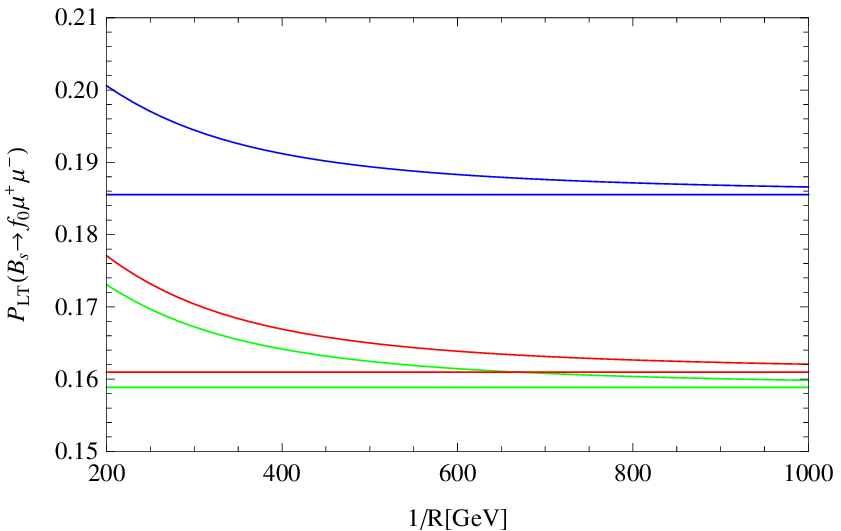}
\includegraphics[width=8cm]{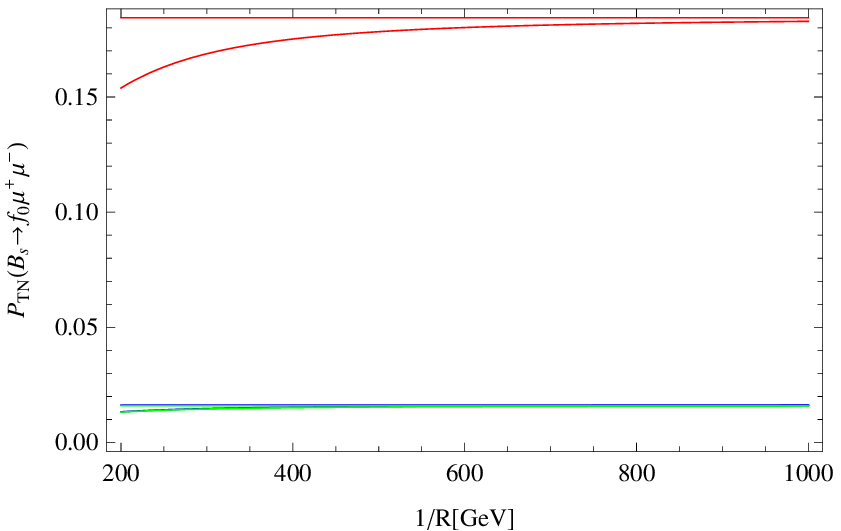}
\end{center}
\caption{The same as FIG. 3 but for $P_{LT}$ and $P_{TN}$.} \label{fig22}
\end{figure}

%
%
\begin{figure}[h!]
\begin{center}
\includegraphics[width=8cm]{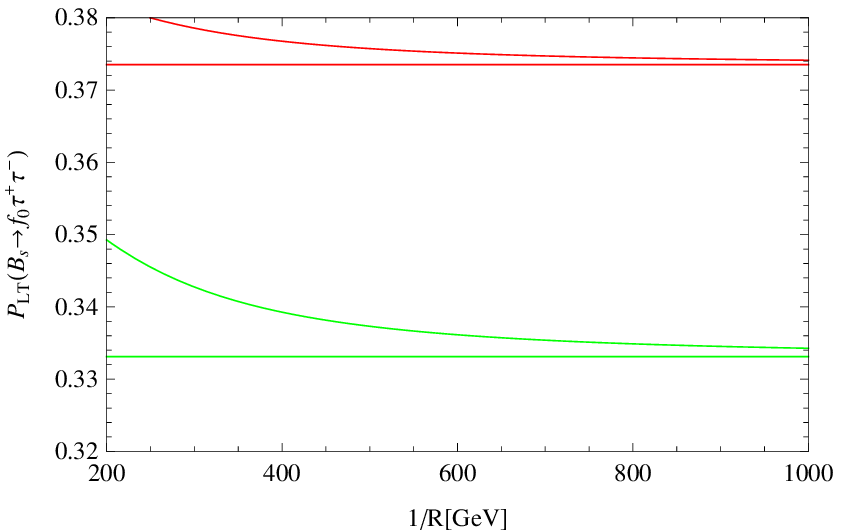}
\includegraphics[width=8cm]{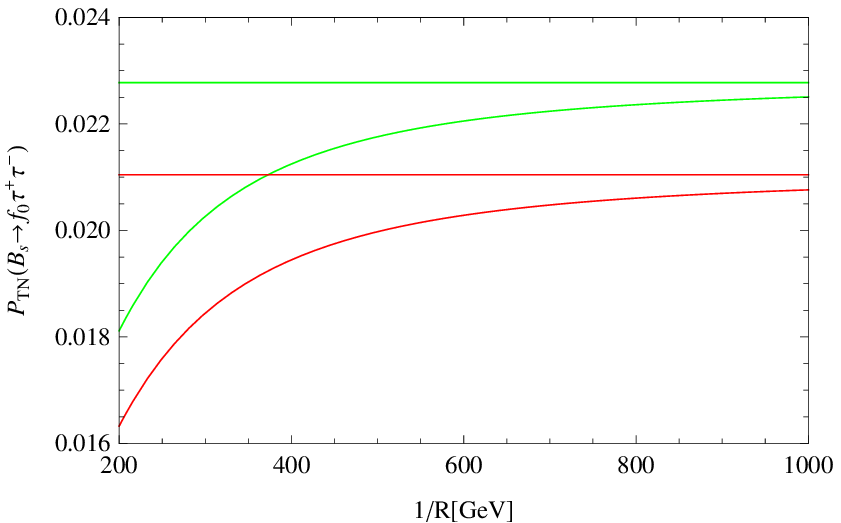}
\end{center}
\caption{The same as FIG. 4 but for $P_{LT}$ and $P_{TN}$.} \label{fig24}
\end{figure}
%
%
\begin{figure}[h!]
\begin{center}
\includegraphics[width=8cm]{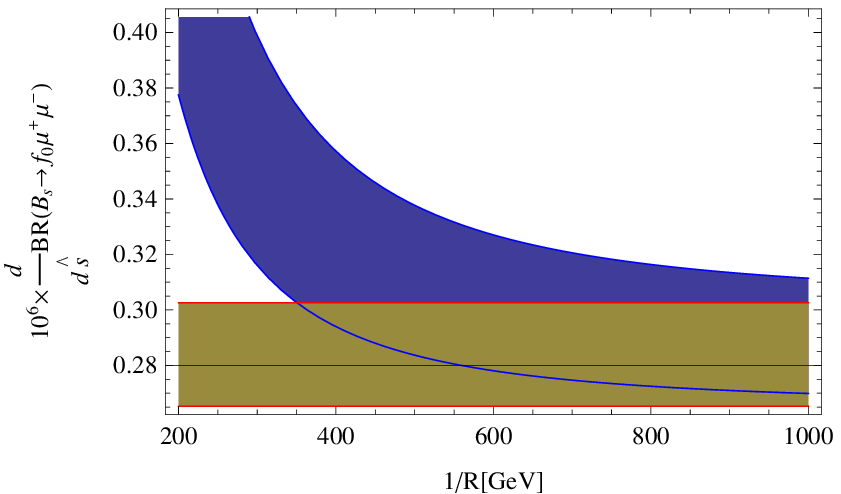}
\end{center}
\caption{Dependence of the branching ratio on the $1/R$ for muon channel  at  $\hat{s}=0.2$ when errors of the form factors are taken into account. The
 straight band shows  result of the  SM and the curve band refers to the ACD model prediction.} \label{fig2}
\end{figure}

\begin{figure}[h!]
\begin{center}
\includegraphics[width=8cm]{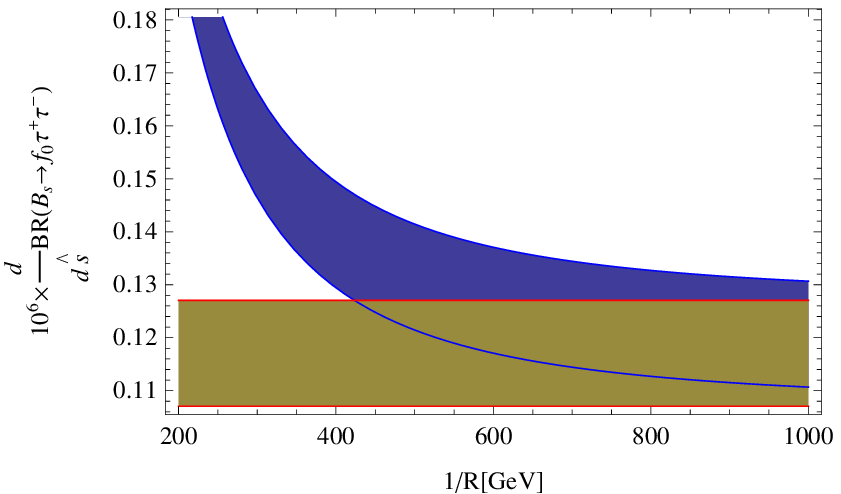}
\end{center}
\caption{Dependence of the branching ratio on the $1/R$ for tau channel  at $\hat{s}=0.6$ when errors of the form factors are taken into account. The
 straight band shows  result of  the SM and the curve band refers to the ACD model prediction.} \label{fig4}
\end{figure}

\begin{figure}[h!]
\begin{center}
\includegraphics[width=8cm]{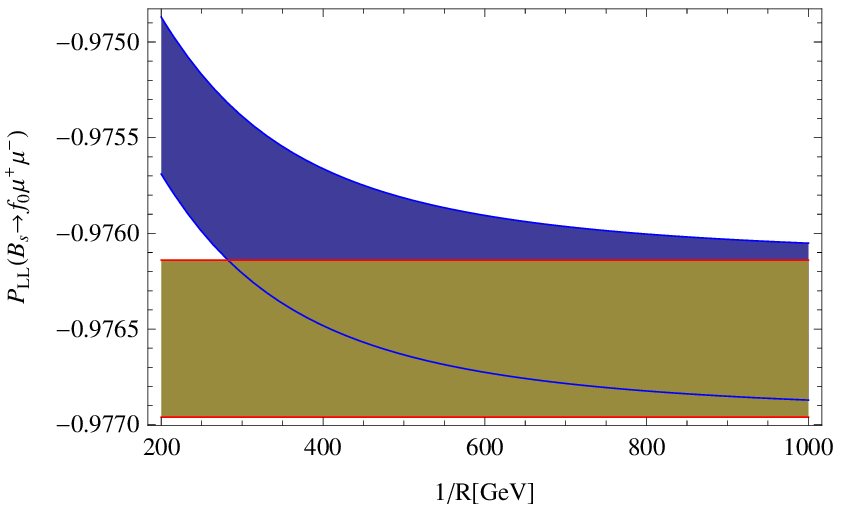}
\includegraphics[width=8cm]{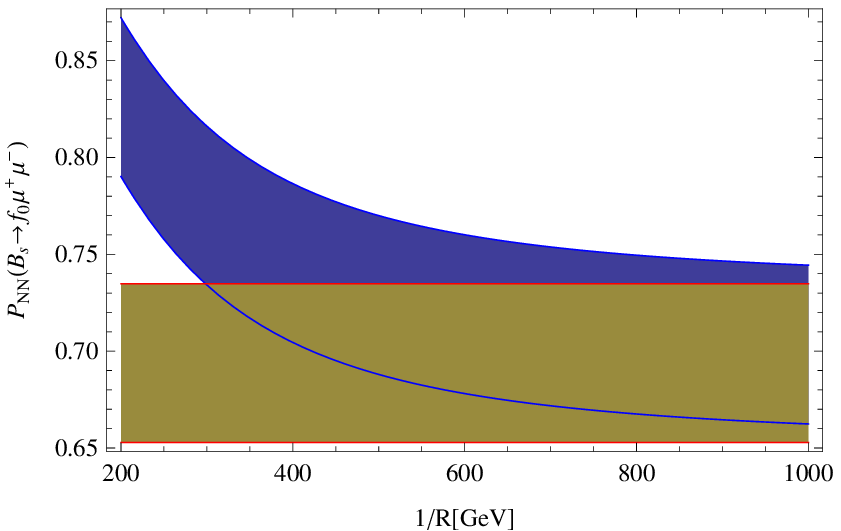}
\end{center}
\caption{Dependence of the $P_{LL}$ and $P_{NN}$ on the $1/R$ for muon channel  at $\hat{s}=0.2$ when errors of the form factors are taken into account. The
 straight bands show  results of  the SM and the curve bands refer to the  the ACD model predictions.} \label{fig66}
\end{figure}

\begin{figure}[h!]
\begin{center}
\includegraphics[width=8cm]{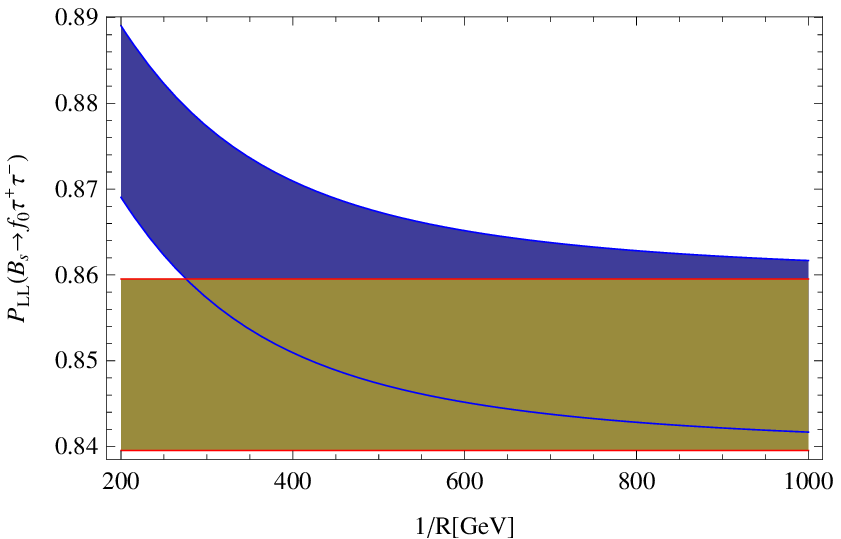}
\includegraphics[width=8cm]{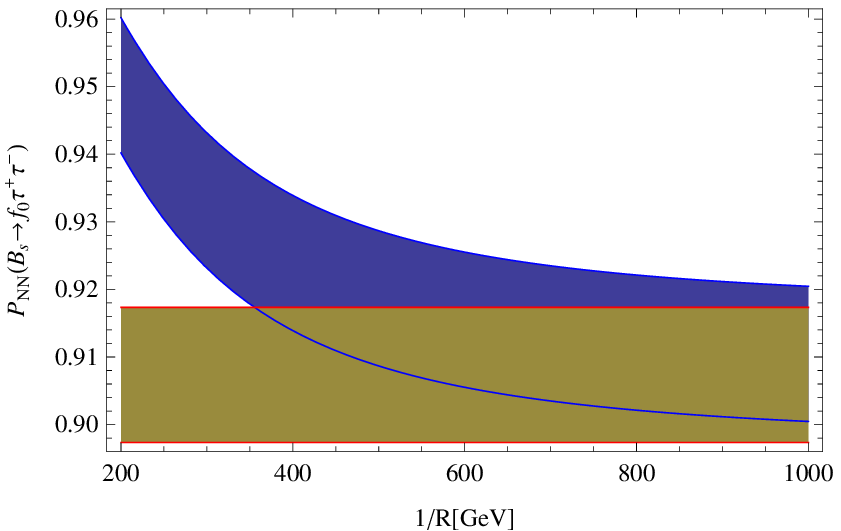}
\end{center}
\caption{Dependence of the $P_{LL}$ and $P_{NN}$ on the $1/R$ for tau channel  at  $\hat{s}=0.6$ when errors of the form factors are taken into account. The
 straight bands show  results of  SM and the curve bands refer to the ACD model predictions.} \label{fig88}
\end{figure}

\begin{figure}[h!]
\begin{center}
\includegraphics[width=8cm]{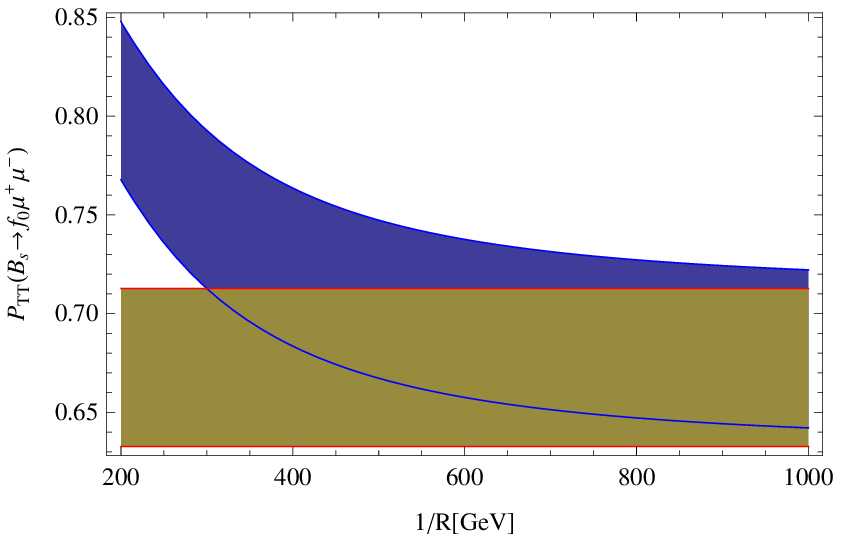}
\includegraphics[width=8cm]{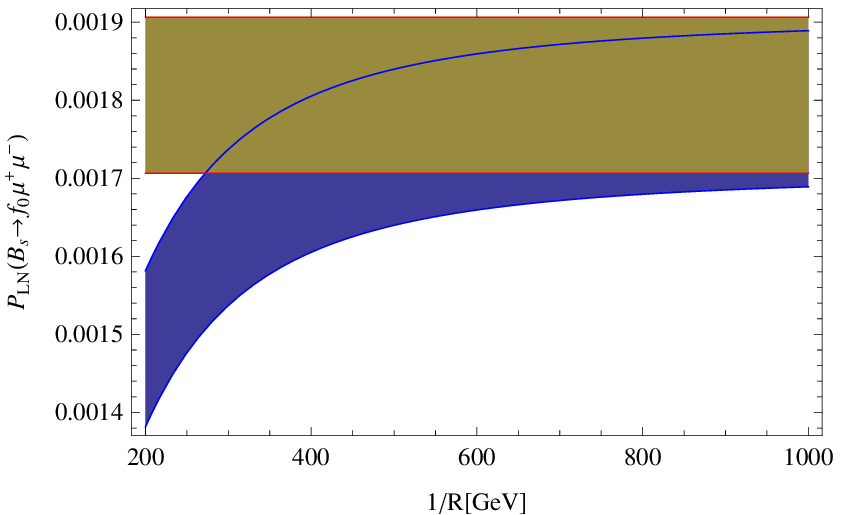}
\end{center}
\caption{The same as FIG. 11 but for $P_{TT}$ and $P_{LN}$.} \label{fig144}
\end{figure}

\begin{figure}[h!]
\begin{center}
\includegraphics[width=8cm]{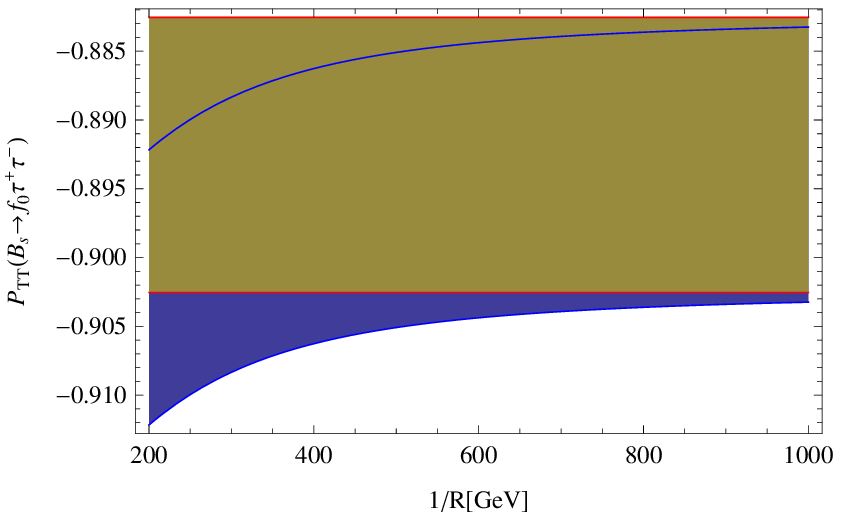}
\includegraphics[width=8cm]{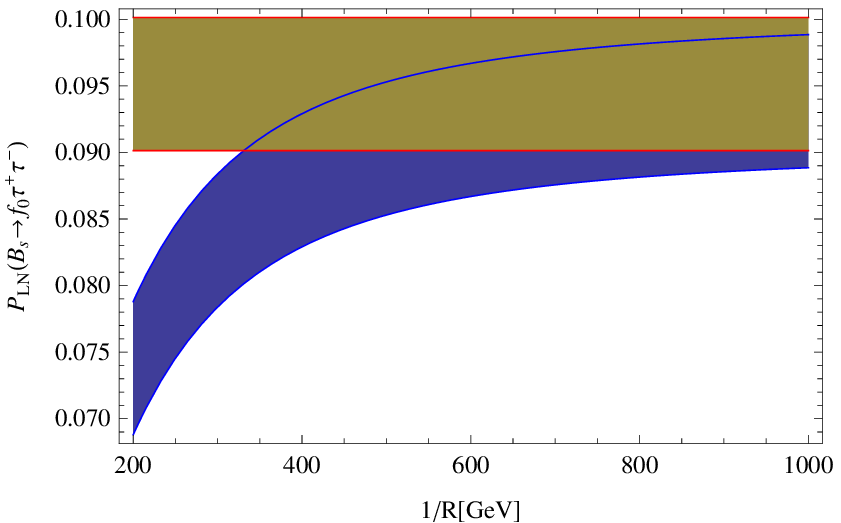}
\end{center}
\caption{The same as FIG. 12 but for $P_{TT}$ and $P_{LN}$.} \label{fig166}
\end{figure}


\begin{figure}[h!]
\begin{center}
\includegraphics[width=8cm]{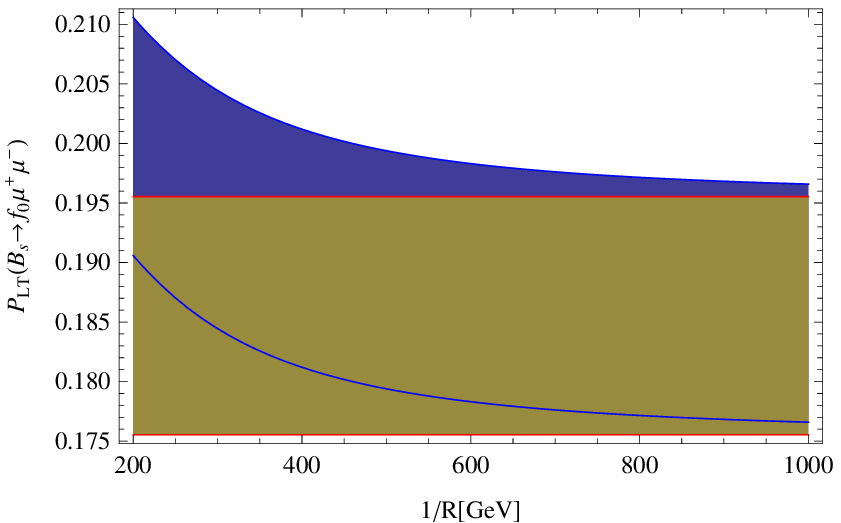}
\includegraphics[width=8cm]{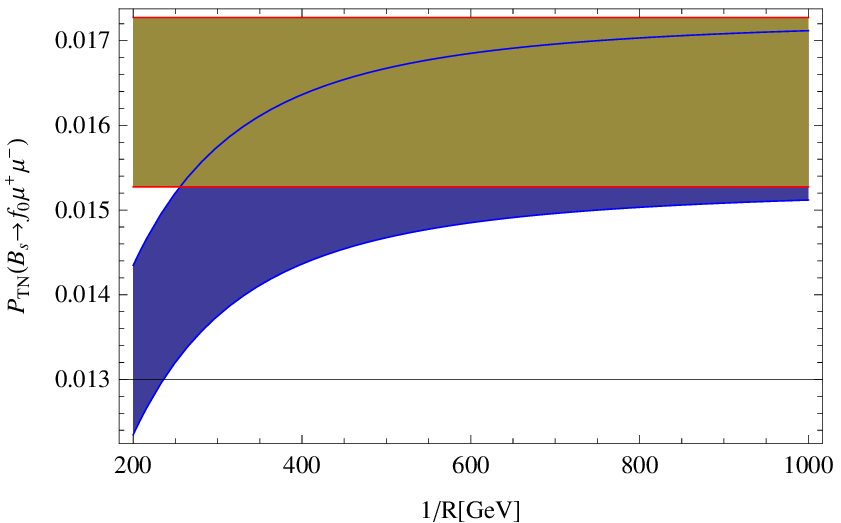}
\end{center}
\caption{The same as FIG. 11 but for $P_{LT}$ and $P_{TN}$.} \label{fig222}
\end{figure}

\begin{figure}[h!]
\begin{center}
\includegraphics[width=8cm]{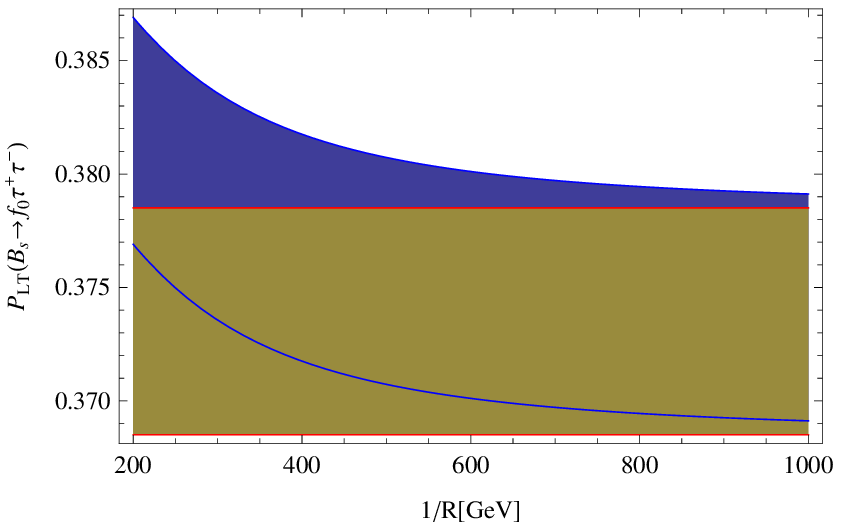}
\includegraphics[width=8cm]{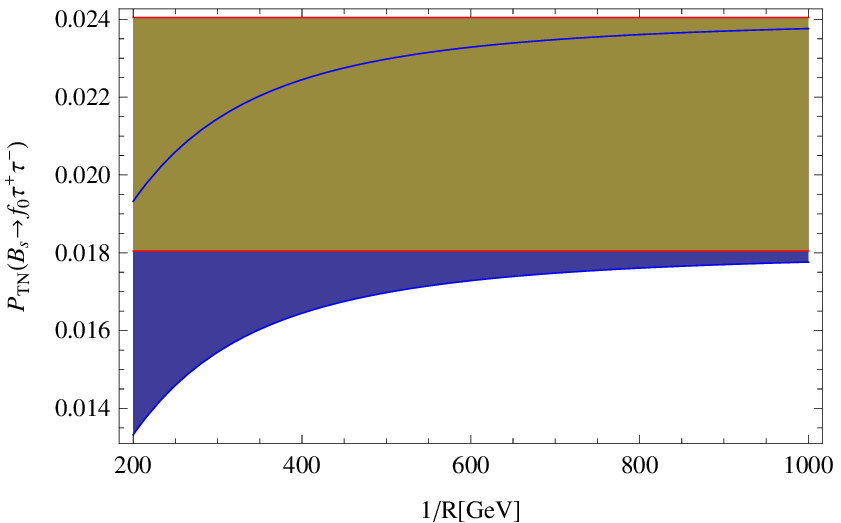}
\end{center}
\caption{The same as FIG. 12 but for $P_{LT}$ and $P_{TN}$.} \label{fig244}
\end{figure}

From these figures, we obtain the following results:
\begin{itemize}
 \item There are considerable discrepancies between the results of the UED and SM predictions at lower values of the compactification factor 
for all observables and both lepton channels. When $1/R$ is increased, the differences between the predictions of two models
become small so that two models have approximately the same predictions at $1/R=1000~GeV$.
\item As it is expected, the branching ratio in $\tau$ channel is small comparing to that of the $\mu$.
\item An increase in the value of $\hat s$ ends up in a decrease in the value of the differential branching ratio.
\item The deviations of the UED results from those of the SM on double lepton polarization asymmetries are small in comparison with the deviation of the differential
 branching ratios from corresponding SM values. However these can not be overlooked.
\item The contributions of KK modes enhance the absolute values of the $P_{LL}$ for $\tau$ as well as the $P_{NN}$, $P_{TT}$ and $P_{LT}$ for both lepton channels,
 but they decrease the
 absolute values of the $P_{LL}$ for $\mu$ and $P_{LN}$ and $P_{TN}$ for both leptons.
\item The $P_{LL}$ for $\mu$ and $P_{TT}$ for $\tau$ have negative signs but the rest of double lepton polarization asymmetries have positive signs.
\end{itemize}

Now, we would like to discuss how the uncertainties of the form factors affect the physical quantities under consideration. For this aim, we plot the aforementioned physical observables
on the compactification factor in figures 9-16 when the uncertainties of the form factors are taken into account. These figures are plotted at  $\hat s=0.2$ and $\hat s=0.6$
for the $\mu$ and $\tau$ channels, respectively. 
 From figures 9 and 10 for the branching fractions,
it is clear that the difference between the UED and SM models predictions exist and can not be killed by uncertainties of the form factors especially at lower 
values of the compactification factor. The figures 11-16 for double lepton polarization
asymmetries also depict that except the $P_{TT}$ and  $P_{TN}$ in $\tau$ channel and $P_{LT}$ in both lepton channels, there are discrepancies between two model predictions at
lower values of the $1/R$ even if the errors of the form factors are encountered.

In conclusion, making use of the related form factors  calculated via light-cone QCD sum rules up to next-to-leading order corrections in $\alpha_s$, we analyzed the sensitivity of the differential
branching ratio and various double lepton polarization asymmetries on the compactification factor of extra dimension for  the $ B_s\rightarrow f_0 (980)\ell^+\ell^- $ transition. 
Our numerical calculations depict  considerable deviations of  the extra dimension model results from the SM predictions. These differences can not be killed by errors of the form factors in the allowed regions
of the compactification parameter previously discussed.  Such discrepancies can be interpreted as 
 signals for existing extra dimensions in nature which can be searched for at hadron colliders.
As a final note it is worth to  estimate  the accessibility to
measure the branching ratio and lepton polarization asymmetries. An
observation of a $3~\sigma$ signal for asymmetry of the order of the
$1\%$ needs about $\sim 10^{12}$ $\bar{B}B$ pairs.  This allow us to measure the  branching ratio and 
the polarization asymmetries  shown in the figures
(1)-(16)  in principle. The order of branching ratios show that the $ B_s\rightarrow f_0 (980)\ell^+\ell^- $ decay channel both for $\mu$ and $\tau$ leptons 
can be detected at LHC. However, as experimentalists say, there are some technical difficulties to measure the lepton polarizations. 
In the case of $\mu$, this lepton should be stopped in order to  measure its polarizations which is not yet possible experimentally.
 For $\tau$ lepton, we should reconstruct then analyze  the decay products of this lepton. In this case, we face with the problem of the efficiency of the reconstruction. 
If these technical difficulties over come, by measuring  the considered double--lepton polarization asymmetries, we can get valuable information about the nature of
 interactions included in the effective Hamiltonian because as the large parts of the uncertainties are  canceled out, the ratio of physical observables 
such as CP, forward--backward asymmetry and single or double--lepton polarization
asymmetries less suffer from the uncertainty of the form factors compared to the branching ratio.

\section{Acknowledgment}
We would like to thank T. M. Aliev for his useful discussions.


\begin{thebibliography}{99}
\bibitem{Colangelo1} P. Colangelo, F. De Fazio, R. Ferrandes and T. N. Pham,
Phys. Rev. D 77, 055019 (2008).
 \bibitem{Colangelo2} M. V. Carlucci, P. Colangelo and F. De Fazio, Phys. Rev. D 80, 055023 (2009).
 \bibitem{acd} T. Appelquist, H. C. Cheng and B. A. Dobrescu, Phys. Rev. D 64, 035002 (2001).
\bibitem{antoniadis1} I. Antoniadis, Phys. Lett. B 246, 377 (1990).
\bibitem{antoniadis2} I. Antoniadis, N. Arkani-Hamed, S. Dimopoulos and G. Dvali, Phys. Lett. B 439, 257 (1998).
\bibitem{arkani}  N. Arkani-Hamed, S. Dimopoulos and G. Dvali, Phys. Lett. B 429, 263 (1998); Phys.
Rev. D 59, 086004 (1999).
\bibitem{Hooper} D. Hooper, S. Profumo, Phys.Rept. 453 (2007) 29.


\bibitem{R7624}
  P. Colangelo, F. De Fazio, R. Ferrandes, T. N. Pham, Phys. Rev. D 73 (2006) 115006.
\bibitem{wangying} Yu-Ming Wang, M. Jamil Aslam, Cai-Dian Lu, Eur. Phys. J. C 59, 847 (2009).
 \bibitem{R7601} T. M. Aliev, M. Savc{\i}, Eur. Phys. J. C 50, 91 (2007).
 \bibitem{fazio} F. De Fazio, Nucl. Phys. Proc. Suppl. 174, 185 (2007).
\bibitem{sirvanli} B. B. Sirvanli, K. Azizi,  Y. Ipekoglu, JHEP 1101, 069 (2011).
\bibitem{kank1}
K. Azizi, N. Kat{\i}rc{\i}, JHEP 1101, 087 (2011).
\bibitem{yee} T. Appelquist, H. U. Yee, Phys. Rev. D 67, 055002 (2003).
\bibitem{bashiry} V. Bashiry, M. Bayar, K. Azizi, Phys. Rev. D 78, 035010 (2008).


\bibitem{aliev} T. M. Aliev, M. Savci, B. B. Sirvanli, Eur. Phys. J. C 52, 375
(2007).
\bibitem{ahmet} I. Ahmed, M. A. Paracha, M. J. Aslam, Eur.
Phys. J. C 54, 591 (2008).

\bibitem{nihan} N. Kat{\i}rc{\i}, K. Azizi, JHEP 1107, 043 (2011).

\bibitem{cog} P. Colangelo, F. De Fazio, R. Ferrandes, T. N. Pham, Phys. Rev. D 74 (2006) 115006.
 \bibitem{Mohanta} R. Mohanta and A. K. Giri, Phys.Rev. D 75 (2007) 035008.
\bibitem{Haisch:2007vb}
U.~Haisch and A.~Weiler,
Phys.\ Rev.\ D  76, 034014 (2007).





\bibitem{Buras:2002ej}
  A.~J.~Buras, M.~Spranger and A.~Weiler,
  Nucl.\ Phys.\ B  660, 225 (2003).
\bibitem{R7623}
  A. J. Buras, A. Poschenrieder, M. Spranger
  Nucl. Phys. B  D 678, 455 (2004).
\bibitem{R7626}
  A. Buras, M. Misiak, M. M\"{u}nz and S. Pokorski,
  Nucl. Phys. B 424, 374 (1994).
\bibitem{R7627}
  M. Misiak,
  Nucl. Phys. B  393, 23 (1993);
  Erratum ibid  B  439, 161 (1995).
\bibitem{R762777}
    B. Buras, M. M\"{u}nz,
  Phys. Rev. D  52, 186 (1995).

\bibitem{ikiuc} K. Agashe, N. G. Deshpande,  G. H. Wu, Phys. Lett. B 514 (2001) 309; B 511 (2001) 85; T. Appelquist,  B. A. Dobrescu, Phys. Lett. B 516 (2001) 85.


\bibitem{Gogoladze:2006br}
I.~Gogoladze and C.~Macesanu,
Phys.\ Rev.\ D  74, 093012 (2006).

\bibitem{Cembranos:2006gt}
J.~A.~R.~Cembranos, J.~L.~Feng and L.~E.~Strigari,
Phys.\ Rev.\ D  75, 036004 (2007).
\bibitem{R8401}
 G. Buchalla, A. J. Buras,  M. E. Lautenbacher,
 Rev. Mod. Phys.  68, 1125 (1996).
\bibitem{cana1} M. Beneke, G. Buchalla, M. Neubert, C.T. Sachrajda,  Eur. Phys. J. C 61 (2009) 439.
\bibitem{cana2} A. Khodjamirian, Th. Mannel, A.A. Pivovarov, Y.-M. Wang, JHEP 1009 (2010) 089. 


\bibitem{colangelo3} P. Colangelo, F. De Fazio, W.  Wang, Phys. Rev.D 81, 074001 (2010).

\bibitem{bas1} T. M. Aliev, V. Bashiry, M. Savci, Eur .Phys. J. C 35 (2004) 197.
\bibitem{bas2} V. Bashiry, S. M. Zeberjad, F. Falahati, K. Azizi, J. Phys. G 35, 065005 (2008) .
\bibitem{Fukae} S. Fukae, C.S. Kim, T. Yoshikawa, Phys. Rev. D 61 (2000) 074015.


   \end{thebibliography}
\end{document}